# The methodology for validation of cross sections in quasi monoenergetic neutron field

Zdeněk Matěj[1], Michal Košťál*[2], Mitja Majerle[3]; Martin Ansorge[3]; Evžen Losa[2], Marek Zmeškal[2]; Martin Schulc[2]; Jan Šimon[2]; Milan Štefánik[3]; Jan Novák[3]; Daniil Koliadko[3], František Cvachovec[4]; Filip Mravec[1]; Václav Přenosil[1]; Václav Zach[3], Tomáš Czakoj[2]; Vojtěch Rypar[2]; Roberto Capote[5]

[1] Masaryk University, Botanická 15, Brno 612 00, Czech Republic
[2] Research Centre Rez, 250 68 Husinec-Rez 130, Czech Republic
[3] UJF Rez, 250 68 Husinec-Rez 130, Czech Republic
[4] University of Defence, Kounicova 65, Brno 612 00, Czech Republic
[5] Nuclear Data Section, International Atomic Energy Agency A-1400 Wien, Austria

Email: Michal.Kostal@cvrez.cz
Telephone: +420266172655

Keywords: IRDFF-II; Cross section validation; $^7$Li(p,n) neutrons; MCNPX

Abstract
Dosimetry cross sections are fundamental quantities essential in determination of neutron fluences in points of interest in technologies under heavy radiation load. The most common is their application to Reactor Pressure Vessel aging management, relating to correct estimation of its residual lifetime. The neutron spectrum in various reactor positions has a similar character as the fission spectrum. Due to this fact, the validation of the neutron dosimetry cross sections in reactor fields, or in a very well-known neutron field of $^{252}$Cf(s.f.) is sufficient for ensuring their validity in estimations of neutron fluxes in reactor physics. With an increasing field of applications, as in neutron dosimetry in accelerator-based fields or space applications, requests occurred on detailed validation in spectra different from fission ones. This paper presents the testing of a new methodology for the use of quasi monoenergetic neutron fields, where different sensitivity allows validations of the cross- section in different energy regions than in the fission spectrum. The exact shape of the neutron spectrum in the tested fields is determined by stilbene spectrometry and corrected to scattering by calculation, where applicable. The total flux is determined from Ni and Al flux monitors. The evaluated neutron flux in the target set of activation foils is used for calculation of theoretical reaction rate, which is compared with the experimental value determined from gamma activity. This kind of comparison can be understood as validation. It's worth noting, this methodology applied to the IRDFF-II library shows satisfactory agreement for selected reactions.

## 1    Introduction

The quasi mono-energetic neutron fields are often used for measurement of differential cross section data, which are input for cross section evaluations [1],[2],[3]. Based on many effects and/or complications during the measurement, the differential data might be burdened by significant uncertainties which even can be up to 30 – 40 %. Sometimes the differential data differ significantly each from other, see Figure 1. Validation of the evaluated cross sections by integral experiments is therefore an essential part of the nuclear data evaluation process. Integral quantity, such as spectral averaged cross section, can usually be measured much more accurately than differential nuclear data, so it is tempting to use this data to refine the evaluation of the cross-section of isotopes.

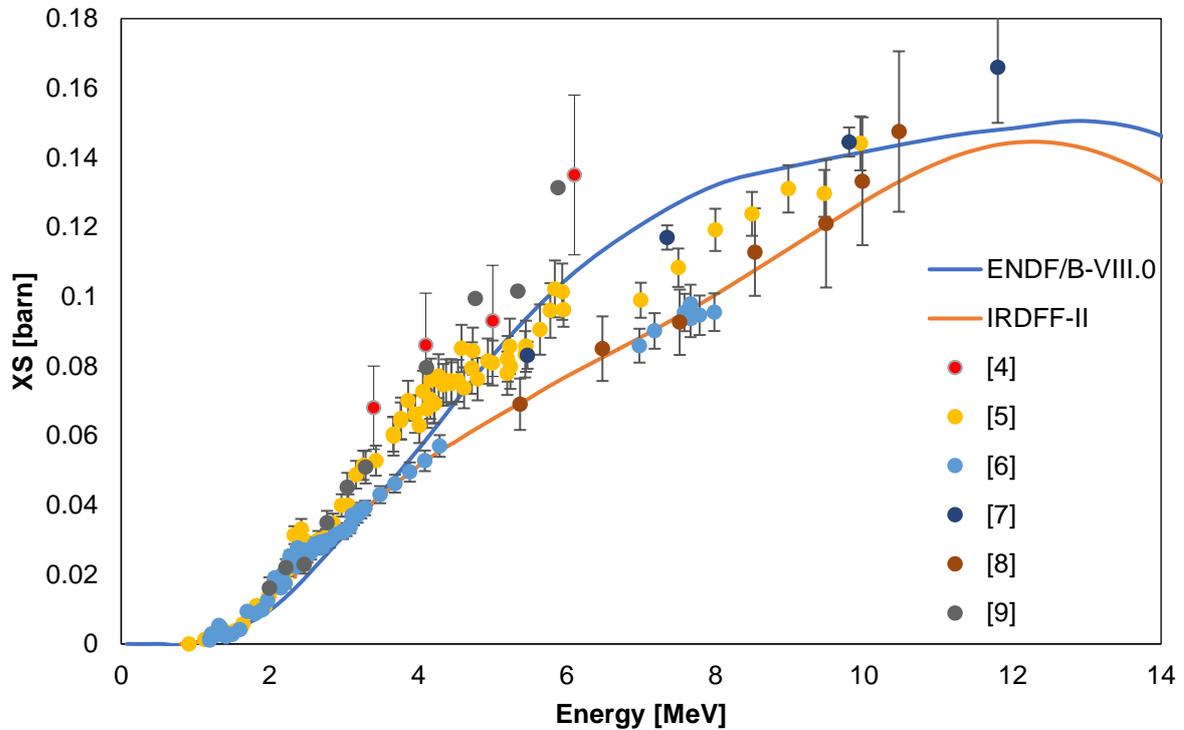

Figure 1: Comparison of various sets of differential data [4-9] and various evaluations in case of $^{47}$Ti(n,p) reaction.

In many cases, the validation of dosimetry cross sections is performed in the fission spectrum produced by $^{252}$Cf(s.f.) or $^{235}$U(n$_{th}$, fiss). In view of the standard reactor dosimetry focused on the determination of the fluences in fission reactors, this approach is satisfactory because the sensitivity of the field used for validation and the field in the application is comparable

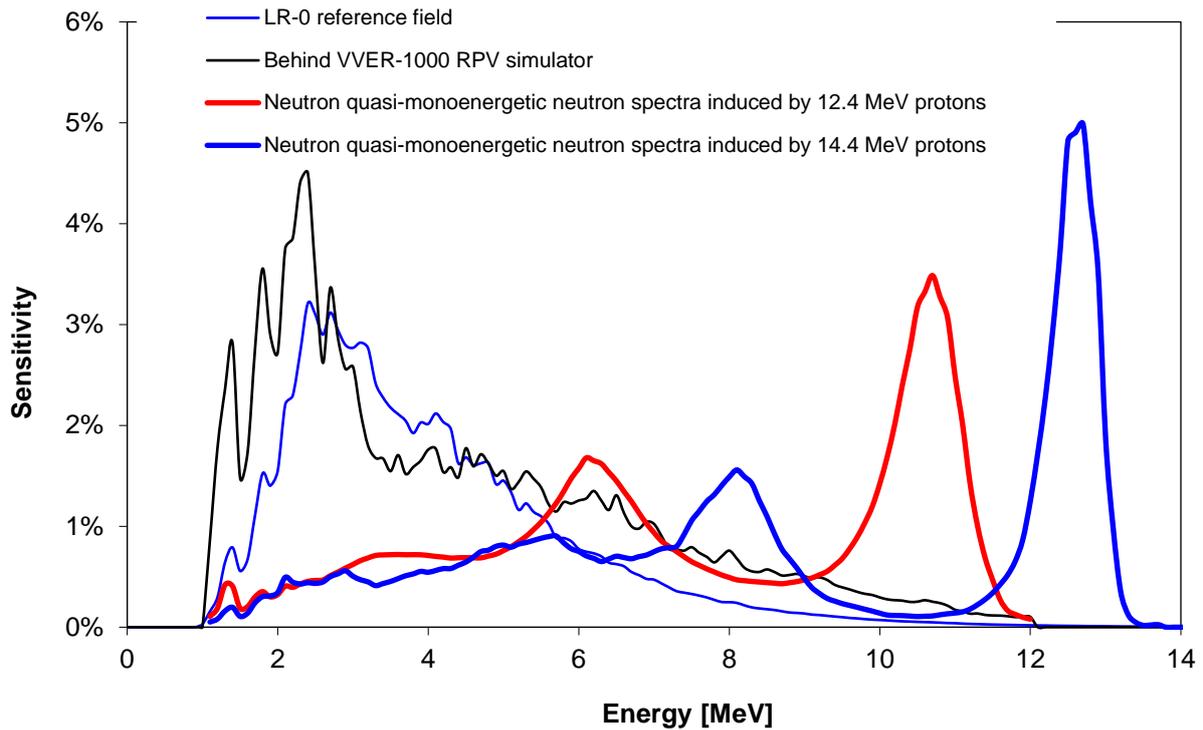

Figure 2.
A question arises when the application field differs from the fission ones. Nowadays, with the increasing number of applications, such as neutron dosimetry in accelerators or space technology, detailed validation in different spectra is requested. This paper deals with the methodology of validation using quasi monoenergetic neutron fields, where different sensitivity allows to validate the cross sections in different energy regions than in fission spectrum-based neutron fields.

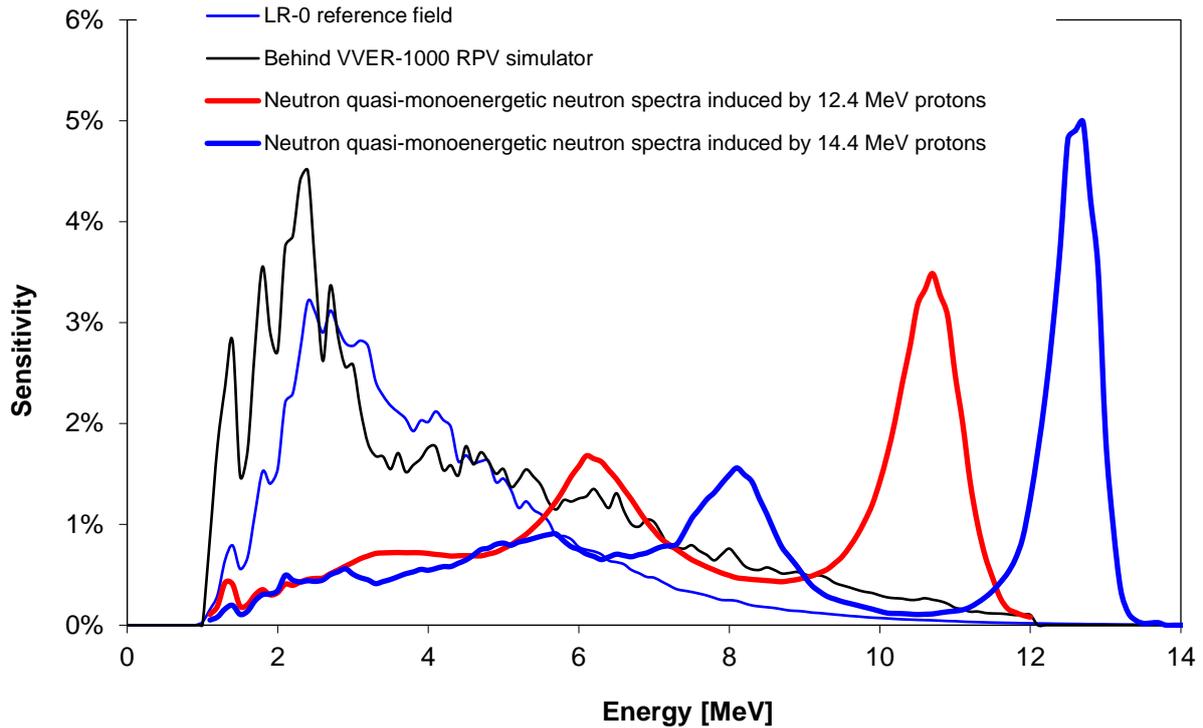

Figure 2: Sensitivities of $^{47}$Ti(n,p) in various neutron fields measured by stilbene detector

## 2 Experimental setup

The irradiation was carried out at the quasi monoenergetic neutron (QM) neutron generator at the Nuclear Physics Institute of the Czech Academy of Sciences (NPI CAS). The generator is based on the $^7$Li(p,n) reaction. It comprises a 0.5 mm thick lithium target ($^{nat}$Li metal) followed by a 1 cm thick carbon slab to stop the protons that remain in the beam after passing the target, see **Chyba! Nenalezen zdroj odkazů.**. The target and slab are ethanol cooled and electrically isolated to allow measurements of a charge brought by incident protons. The beam of protons, accelerated by the U120M cyclotron and directed to the target, produces a QM neutrons. The proton energy can be set between 10-35 MeV. The design of the generator allows the extraction of a lithium target after irradiation (for γ-measurements). The cyclotron radio frequency (RF) repetition period of 40-60 ns allows time-of-flight (TOF) measurements of neutron spectra. Further details can be found in [10].

Two proton irradiations of 12.4 and 14.4 MeV were performed. The cyclotron radiofrequency during the first irradiation was 16.295 MHz (resulting in a repetition rate of 61.4 ns). The beam spot was described by two Gaussians in X/Y planes with the FWHM of 3 and 5 mm. During the second irradiation, the RF was 17.124 MHz (repetition rate 58.4 ns), and the beam spot dimensions were Gaussians with FWHM of 3 and 4 mm.

The neutrons produced in the lithium target pass a carbon slab and 6 mm of ethanol coolant, pass 4 cm to the exit window (0.5 mm Al), and through the air to the activation detectors located 8.6 mm behind the Li target and further to the scintillation detectors at the distances ~ 4 m from the front of the Li target.

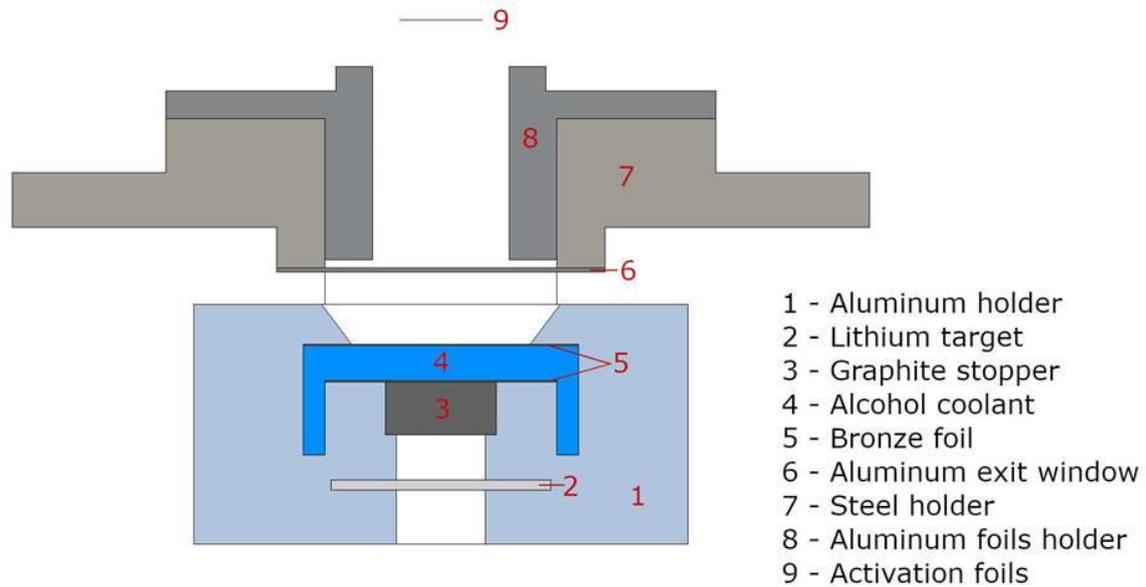

1 - Aluminum holder
2 - Lithium target
3 - Graphite stopper
4 - Alcohol coolant
5 - Bronze foil
6 - Aluminum exit window
7 - Steel holder
8 - Aluminum foils holder
9 - Activation foils

Figure 3: Target geometry

Neutron flux density in the target set of activation foils was determined by a set of monitoring foils which were in one stack together with the validated foils. The foils in the target were in the form of a stack 1.5 cm in diameter and 2.15 mm thick in the first irradiation and 3.82 mm in the second case. Flux monitors made of pure $^{nat.}$Ni with a diameter of 1.5 cm and a thickness of 0.01 mm were placed in 0.2 mm increment or after each thick foil. In the first experiment, the following stack of monitors in direction from Li foil proton-neutron converter with a given thickness was used: Fe 0.1 mm, Au 0.05 mm, Mg 0.05 mm, Ti 0.2 mm, Co 0.25 mm, Al 0.25 mm, V 0.075 mm, Fe 0.1 mm, and Ni 1 mm. The second target was designed in similar geometry, only the second Fe foil was replaced by pure $^{54}$Fe foil and after the Ni foil, PTFE ($CF_2$) and Mn foils were placed. The composition was as following: Fe 0.1 mm, Au 0.05 mm, Mg 0.05 mm, Ti 0.2 mm, Co 0.25 mm, Al 0.25 mm, V 0.075 mm, $^{54}$Fe 0.05 mm, Ni 1 mm, $CF_2$ 1 mm, and Mn 0.7 mm.

## 3    Experimental and calculation methods

A set of neutron spectra in the energy range 0.9 MeV to 14.5 MeV was measured by the proton-recoil method using a Stilbene scintillator (10×10 mm) with neutron and gamma pulse shape discrimination. The independent measurement with TOF method was performed as well.

The experimental reaction rates were derived from the gamma activities of irradiated target, the set of activation foils and lithium target. They were measured using a well-defined HPGe detector with an experimentally validated efficiency curve. The calculations of response functions, efficiency and neutron production in Li target were carried out using the MCNP6.2 code [11]. The neutron production was simulated also using GEANT4-10.7.1 [23].

### 3.1    Stilbene measurement by deconvolution

The measurements of neutron spectra in narrow groups were performed using two-parameter spectrometric system NGA-01 [12], [13]. This fully digitized device can process up to 500 000 impulse responses per second. Used system works with high-speed ADC converters with 500 MS / s (alternatively 1 GS / s) and a size of 12 bits. Field Programmable Gate Area (FPGA) with advanced digital filters and PSD algorithms ensures lossless data processing. This achieves the processing of each impulse response from the detector without the dead time of the digital part. The processing parameters are set depending on the used configuration of the detector with a preamplifier. A high voltage source is also integrated directly into the system to eliminate external noise.

The used scintillation crystal, cylindrical stilbene of diameter 10 mm and height 10 mm, is sensitive to both neutrons and gammas. Separation between both signals can be realized by pulse shape discrimination employing different shape of the signal caused by neutrons or gammas [14].

The pulse shape discrimination (D) is realized within the Field Programable Gate Area by an integration algorithm, the principle of which consists in comparing the area limited by the part of a trailing edge of the measured response ($Q_1$) with the area limited by the whole response ($Q_2$). The areas $Q_1$ and $Q_2$ as the integrals in time are expressed in Eq. 1 and their illustration is shown in Figure 4. The time offset for calculating the area of the integral $t_2$ depends on the time constant of the apparatus and the use of a scintillator from the maximum amplitude ($t_1$). For stilbene and a 50 ohm working resistor at the anode of the photomultiplier, it is usually in the range of 5-16 ns.

The obtained recoiled protons are then subjected to deconvolution by the Maximum Likelihood Estimation [15]. The photon yields were derived from [16].

The robustness of used methodology was tested in various reactor fields [17] as well as accelerator fields [18]. Energy calibration was tested in a QM field [19].

$$Q_1 = \int_{t_2}^{t_3} i(t)dt, \quad Q_2 = \int_{t_0}^{t_3} i(t)dt, \quad D = \frac{Q_1}{Q_2} \tag{1}$$

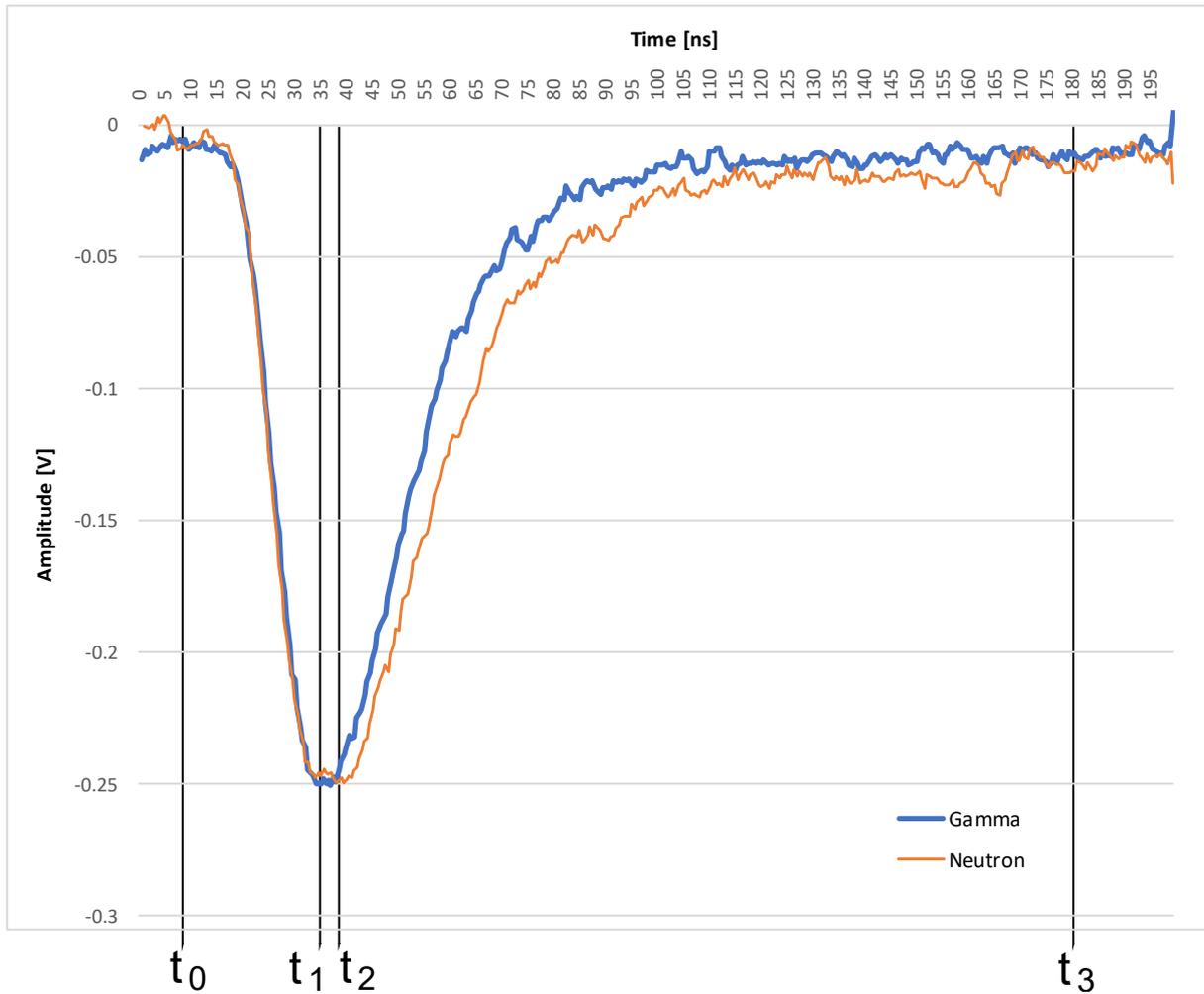

Figure 4: Comparison of similar amplitude neutron and gamma signal with indicated with pulse start ($t_0$), maximum amplitude ($t_1$), offset ($t_2$) and pulse end ($t_3$)

### 3.2 TOF measurement

A NE213 (2"×2") and stilbene (10×10 mm) scintillators were placed on the beam axis at distances from 2 m to 4.5 m from the target output. Pulse shapes from the scintillator were recorded simultaneously with a cyclotron radiofrequency (RF) signal with a 2GS/s and a 10bit digitizer. For each recorded pulse, the time phase related to the RF signal and the signal surface integrals in two-time windows were determined. The pulse shape discrimination method described above was used to separate the signals from neutrons and gammas.

The beginning of each pulse was determined with reference to the RF signal. The time distribution of gammas shows the time distribution of the proton beam and determines the time resolution of the experiment. The time distribution of the proton beam was complex and consisted of several smaller peaks for the irradiation with 12.4 MeV protons. For the irradiation with the 14.4 MeV protons, the time resolution of the proton beam was tuned to one narrow peak with the FWHM of 2 ns.

The amplitude of the neutron pulses versus the determined time phase to the RF signal for the irradiation with 14.4 MeV protons is shown in Figure 5. The repetitive pattern is seen with a period of 58.4 ns, which corresponds to the cyclotron RF of 17.124MHz. Gammas travel13.5 ns (stilbene at a distance of 4.041 m) to the scintillator and the X-axis is shifted to bring the main gamma peak to the right time (13.5, 71.9, 130.3, ns). The most energetic neutrons need 82.6 ns to reach the same distance. The upper black line corresponds to the neutron energy calculated from their time of flight and is converted to the electron light output of recoiled hydrogen nucleus. The dashed black line is the same quantity but shifted one frame back in time.

To obtain the final spectral points the pulses registered in the interval between the red lines from the Figure 5 were summed into a time histogram, which was converted to an energy histogram using the relativistic Time-Of-Flight to energy formalism and divided by the calculated scintillator efficiency. Using only the upper part of the scintillator response (dynamic threshold method) allow the separation of the first frame neutrons from the slower ones. Due to frame overlap, it was not possible to measure the neutron spectrum below the neutron energy of 4 MeV. The resulting energy spectra are shown in Figure 9.

The scintillator response matrix was calculated using the MCNP6.2 [11] code with the JEFF-3.3 [36] database for both scintillators and is shown in Figure 6. The response corresponding to the interval between the red lines was summed to obtain the efficiency of neutron detection - the probability that the neutron is registered in the interval between the red lines.

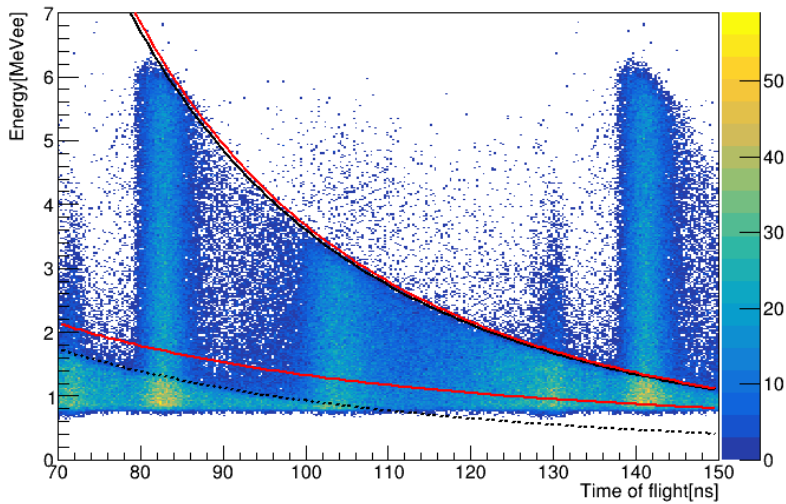

Figure 5: Amplitude of pulses versus their phase to the RF signal.

The repetitive pattern on the scale 58.4 ns is seen. The time frames are shifted and copied for ease of understanding.

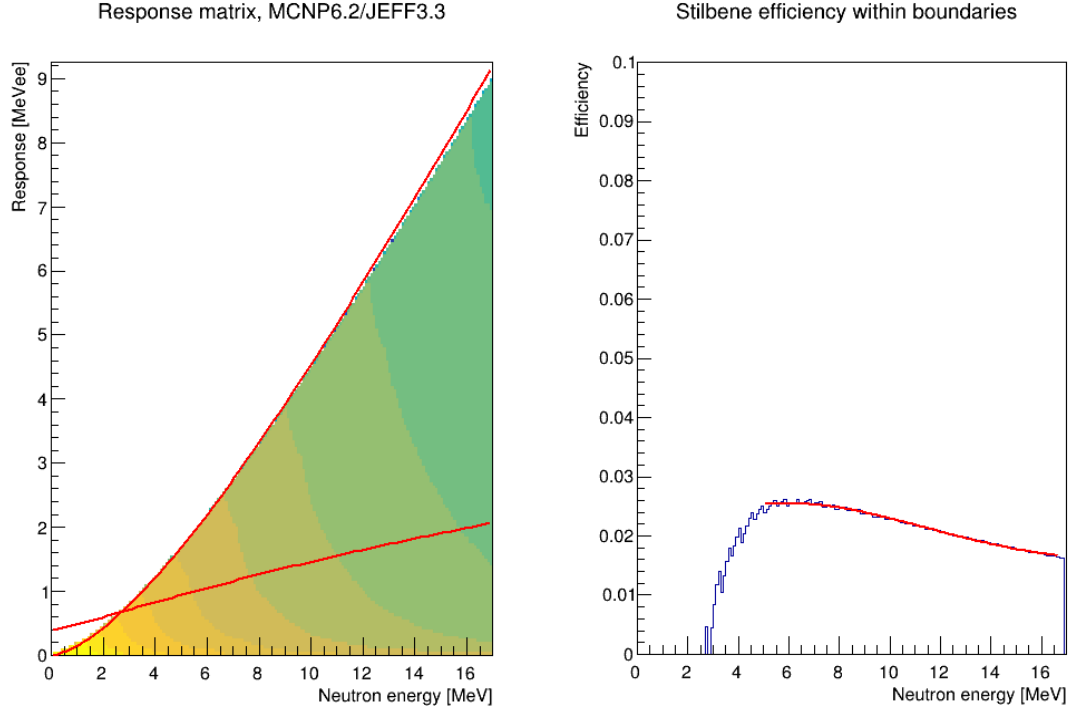

Figure 6: Response of the 10×10 mm stilbene scintillator to neutron energies at a range of 0-17 MeV calculated with a MCNP6.2 code and JEFF 3.3 library. The calculated efficiency in the interval between the red lines is shown on the right.

.

### 3.3   Reaction rates measurement

The experimental reaction rates were derived from gamma activities (Equations 2 - 4) and knowledge of irradiation run (see Figure 7) using Equation 3. Activities were determined by gamma spectrometry of activation products using semiconductor spectrometry with a well determined efficiency curve  [20]. Such approach allows the calculation of efficiency as well as coincidence summing corrections for arbitrary geometry of gamma emitting sources [21]. This is very important especially for any experiment with a low activation rate, where for obtaining sufficient Net Peak Areas the sample have to be measured in close geometry, namely End Cap geometry. Due to various materials and even thicknesses of foils, the simulated efficiency curve is the only realistic approach to evaluation. The uncertainty of HPGe coming from uncertainties of material and geometrical parameters was determined to be 1.8 % for measurements when the gamma source was placed on the HPGe cap [20].

The evaluated reactions were as follows: $^{58}$Ni(n,p), $^{58}$Ni(n,x)$^{57}$Co, $^{56}$Fe(n,p), $^{54}$Fe(n,p), $^{54}$Fe(n,α), $^{197}$Au(n,2n), $^{24}$Mg(n,p), $^{46}$Ti(n,p), $^{47}$Ti(n,p), $^{48}$Ti(n,p), $^{27}$Al(n,α), $^{59}$Co(n,α), $^{59}$Co(n,p),$^{59}$Co(n,2n), $^{51}$V(n,α), $^{19}$F(n,2n), and $^{55}$Mn(n,2n).

The $^7$Be activity of the irradiated lithium targets were measured with the HPGe at sample-detector distance with the total uncertainty of 2-3 %.

The detailed summary of the activation dosimeters used is given in Table 1. To ensure the precise position of thin foils, all dosimeters with a foil thickness less than 0.5 mm were fixed in the EG3 etalon plastic capsule. This means that all foils except Ni, PTFE, and Mn were in the EG3 holder, which ensures that the foil is 1.4 mm above the end cap. This approach slightly decreases the efficiency of this geometry but ensures repeatability of measurement.

Short lived isotopes and $^{48}$Sc formed in vanadium and titanium dosimeters were measured immediately after irradiation. The other foils were measured after decay into ground state. This is issue for example in case of cobalt isotope $^{58}$Co and reactions $^{59}$Co(n,2n) and $^{58}$Ni(n,p). In the case of the experiment with protons irradiation of 14.4 MeV and $^{58}$Co originating in $^{59}$Co(n,2n) reaction, about 69 % of $^{58}$Co is originating in the metastable state, in the case of $^{58}$Ni(n,p) the share of $^{58}$Co originating in metastable state is about 19.4 %.

Due to low share of neutrons in region above 13 MeV and low cross-sections of (n,n+p) reactions for titanium isotopes, contribution of these reactions can be neglected. Namely, in case of $^{47}$Sc the contribution from $^{48}$Ti(n,n+p) is about 0.03 %, in case of $^{48}$Sc the contribution from $^{48}$Ti(n,n+p) is about 0.001 %.

$$q\left(\overline{P}\right) = A_{End} \times \left( \frac{A\left(\overline{P}\right)}{A_{Sat}\left(\overline{P}\right)} \right)^{-1} \qquad (2)$$

$$\frac{A\left(\overline{P}\right)}{A_{Sat}\left(\overline{P}\right)} = \sum_i P_{rel}^i \times \left(1 - e^{-\lambda.T_{Irr}^i}\right) \times e^{-\lambda.T_{end}^i} \qquad (3)$$

$$A_{End.} = NPA\left(T_{Meas.}\right) \times \frac{\lambda}{\varepsilon \times \eta \times N} \times \frac{1}{\left(1 - e^{-\lambda.T_{Meas.}}\right)} \times \frac{1}{e^{-\lambda.\Delta T}} \times k_{CSCF} \qquad (4)$$

Where:

$q$ ; is the reaction rate of activation during irradiation batch

$\lambda$ ; is the decay constant of the radioisotope considered;

$T_{Meas.}$ ; is a time of measurement by the HPGe;

$\Delta T$ ; is the time between the end of irradiation and the start of the HPGe measurement;

$NPA(T_m)$ ; is the Net Peak Area (the measured number of counts);

$\varepsilon$ ; is the gamma branching ratio;

$\eta$ ; is the detector efficiency (it's being determined via MCNP6.2 calculation);

$N$ ; is the number of target isotope nuclei;

$k_{CSCF}$ is coincidence summing correction factor

$T_{Irr.}$ ; is the end of the irradiation period

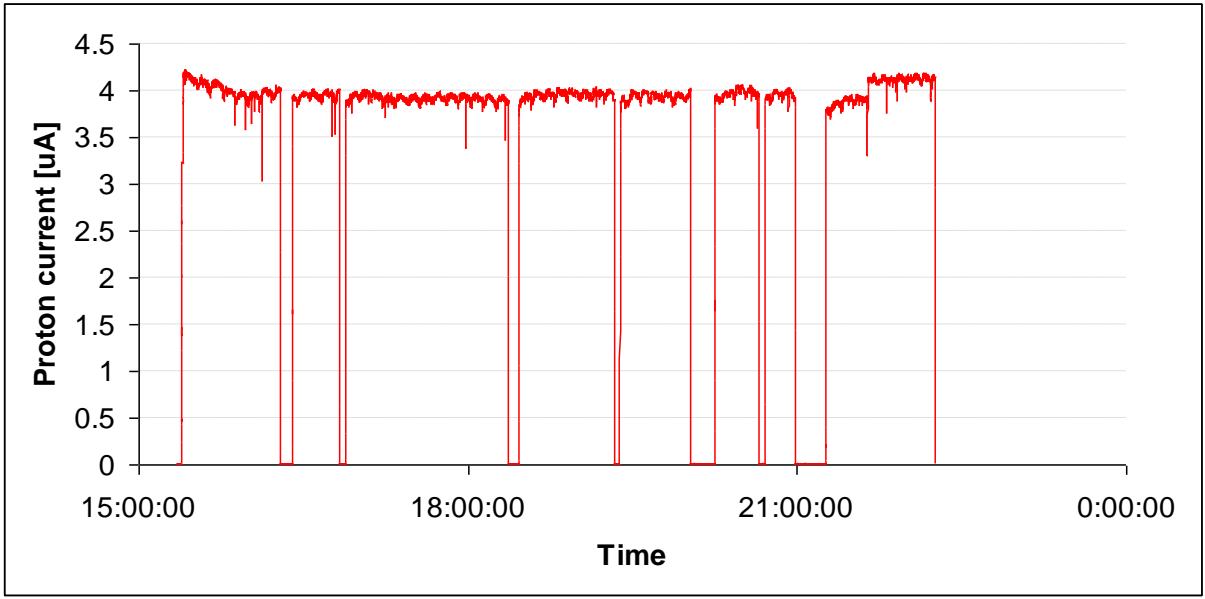

Figure 7: Proton current during irradiation experiment in 12.4 MeV beam

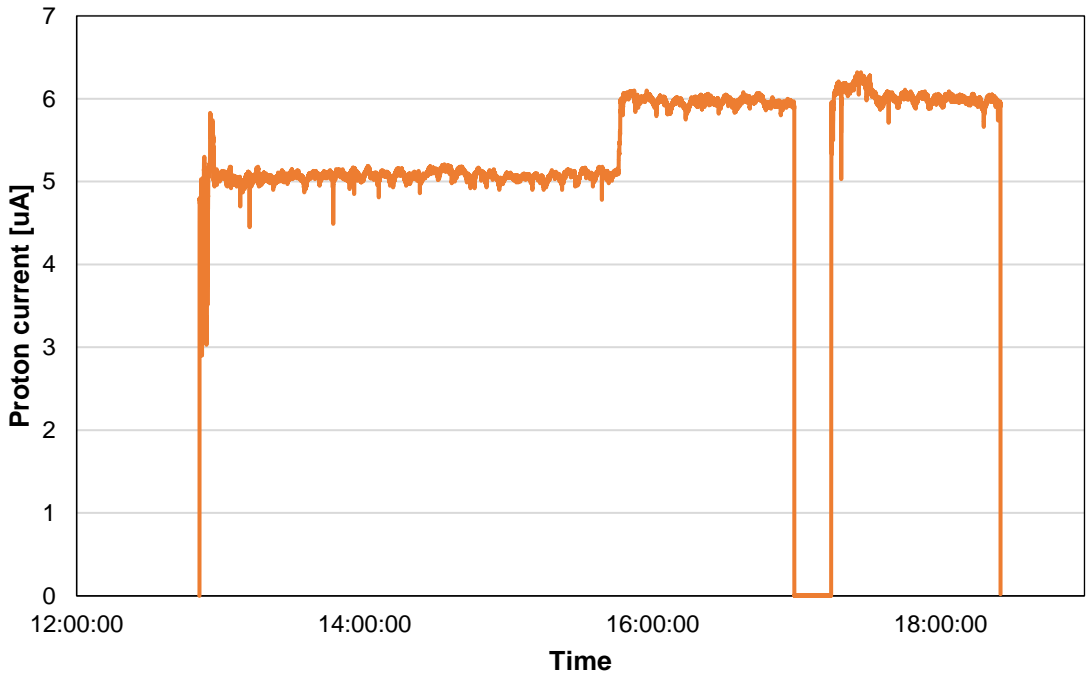

Figure 8: Proton current during irradiation experiment in 14.4 MeV beam

Table 1: Used activation detectors and lithium target

| Foil | Thickness | Reaction | Peak [keV] | Efficiency | CSCF |
|------|-----------|----------|------------|------------|------|
| Ni (D=1.5 cm) | 0.01 mm | $^{58}$Ni(n,p) | 810.8 | 4.52E-2 | 0.938 |
| Fe (D=1.5 cm) | 0.1 mm | $^{56}$Fe(n,p) | 846.8 | 4.34E-2 | 0.939 |
| | | | 1810.7 | 2.31E-2 | 0.815 |
| Fe (D=1.5 cm) | 0.2 mm | $^{54}$Fe(n,p) | 834.8 | 4.32E-2 | 1.000 |
| $^{54}$Fe (D=1.5 cm) | 0.05 mm | $^{54}$Fe(n,p) | 834.8 | 4.40E-2 | 1.000 |
| | | $^{54}$Fe(n,$\alpha$) | 320.1 | 1.01E-1 | 1.000 |
| Au (D=1.5 cm) | 0.05 mm | $^{197}$Au(n,2n) | 333.0 | 9.64E-2 | 0.807 |
| | | | 355.0 | 9.10E-2 | 0.987 |
| Mg (D=1.5 cm) | 0.05 mm | $^{24}$Mg(n,p) | 1368.6 | 2.94E-2 | 0.863 |
| Ti (D=1.5 cm) | 0.2 mm | $^{47}$Ti(n,p) | 159.4 | 1.64E-1 | 1.000 |
| | | $^{46}$Ti(n,p) | 889.3 | 4.16E-2 | 0.829 |
| | | $^{48}$Ti(n,p) | 983.5 | 3.84E-2 | 0.666 |
| | | | 1037.5 | 3.67E-2 | 0.659 |
| Co (D=1.5 cm) | 0.25 mm | $^{59}$Co(n,$\alpha$) | 846.8 | 4.30E-2 | 0.939 |
| | | | 1810.7 | 2.29E-2 | 0.814 |
| | | $^{59}$Co(n,p) | 1099.2 | 3.48E-2 | 1.000 |
| | | $^{59}$Co(n,2n) | 810.8 | 4.42E-02 | 0.937 |
| Al (D=1.5 cm) | 0.25 mm | $^{27}$Al(n,$\alpha$) | 1368.6 | 2.92E-2 | 0.863 |
| Ni (D=1.5 cm) | 1 mm | $^{58}$Ni(n,p) | 810.8 | 4.64E-2 | 0.933 |
| | | $^{60}$Ni(n,p) | 1173.0 | 3.45E-2 | 0.823 |
| | | | 1332.5 | 3.11E-2 | 0.817 |
| | | $^{58}$Ni(n,x)$^{57}$Co | 122.0 | 1.48E-1 | 1.000 |
| V (1×1 cm) | 0.075 mm | $^{51}$V(n,$\alpha$) | 983.5 | 3.76E-2 | 0.666 |
| | | | 1037.5 | 3.60E-2 | 0.659 |
| CF$_2$ (D=1.5 cm) | 1 mm | $^{18}$N(n,2n) | 511.0 | 7.08E-02 | 1.000 |
| Mn (9 × 7.66) | 0.7 mm | $^{55}$Mn(n,2n) | 834.8 | 4.71E-02 | 1.000 |
| Li target | 0.5 mm | $^{7}$Li(p,n)$^{7}$Be | 477.6 | 1.044E-3 | 1.000 |

### 3.4 Calculation methods

The neutron field produced by the 12.4 MeV and 14.4 MeV protons bombarding a 0.5 mm thick lithium target was simulated using different codes and libraries. The only libraries evaluated with the proper description of the neutron production by the p+$^{7}$Li reaction are LA150H and JENDL4/HE. Both libraries were used with the MCNP6 [22] codes to simulate the neutron spectrum for our experimental setup at the place of the scintillator detectors.

The second approach was based on the Geant4 [23] code with the costum hardcoded libraries for better approximation of neutron production and transport. The hadronic models with the binary cascade [26] through G4HadronPhysicsQGSP_BIC_HP and the charge exchange process through G4ChargeExchange were used for this simulation. This combination was used in **Chyba! Nenalezen zdroj odkazů.** where it was used to simulate neutrons from the $^{9}$Be(p,n)$^{9}$B reaction with 20 to 35 MeV protons. It was shown that the binary cascade is capable of simulation of continuum neutrons from the three-body break-up reaction, whereas the charge exchange model allows to simulate neutrons in peaks from direct (p,n) reactions.

To simulate the neutron spectrum at the position of the activation foils, defined source approach was used. The neutron spectrum at the position of the activation foils (8.6 cm from lithium target) is affected by the neutron scattering in the target station material and does not correspond to the spectrum measured by the stilbene scaled down to the distance of 8.6 cm considering just the fraction of the detection solid angles and air attenuation factor). The neutron spectrum measured by stilbene was at first approximated by the sum of separate neutron sources as described below:

The starting point of the neutrons was in the volume of the lithium target, inside the beam spot described by two perpendicular gaussians with FWHM corresponding to experimentally measured beam spot.

The background neutrons - 12 neutron sources with the energies of 1, 2, 3, ... 10, 11, and 12 MeV (the energies were smeared with the gaussian of 0.5 MeV FWHM) with the isotropic angular distribution (in the COM system) were defined to cover the neutrons produced by the compound nucleus mechanism.

The peak neutrons - 2 neutron sources with the energies of 10.4 MeV and 12.4 MeV corresponding to the measured neutron peak energies (the energy of the peaks was smeared with the gaussian of 0.5 MeV FWHM) with the angular distribution taken from [28] (for the closest energy of 15.1 MeV) were defined to cover the peak neutrons produced by the direct reaction mechanism. The angular distribution of neutron peaks at the energies of 6 MeV and 8 MeV corresponding to the second excited state of the $^7$Be was measured to be merely isotropic [29] and they were covered by background neutrons.

The transport of neutrons from the described neutron sources to the position of activation foils and stilbene scintillator detector was simulated using MCNP6 code using ENDFB-VIII.0 data library. The separate contributions were summed with weights adjusted so that the resulting neutron spectrum at the position of the scintillator corresponded to the experimentally determined spectrum. The spectrum obtained with the same weights at the distance of 8.6 cm from the lithium target was then compared to the scaled down spectrum from the position of the scintillator detector. See Figure 12. The resulting ratios represent the coefficients which were used to transform the measured neutron spectra from the position of scintillator detectors to the position of activation foils.

## 4  Results

### 4.1  Measured differential neutron spectra

The neutron spectrum in distance 403.85 cm from $^{nat.}$Li target for both proton energies of 12.4 and 14.4 MeV was measured using $10 \times 10$ mm stilbene detector. For comparison, the spectra were evaluated using standard deconvolution method [15] and by TOF. The comparison of both is plotted for 12.4 MeV protons interacting with $^{Nnat.}$Li target in Figure 9, while for 14.4 MeV protons are plotted in Figure 10.

Both experimentally determined spectra are in good agreement. In deconvolution methodology, the uncertainties in such type of field are about $5 - 10$ %. In wide groups, covering the peak area, the uncertainties are significantly smaller being $2 - 3$ %.

The neutron spectra for the first irradiation with 12.4 MeV protons are listed in Table 4, for 14.4 MeV protons in Table 5.

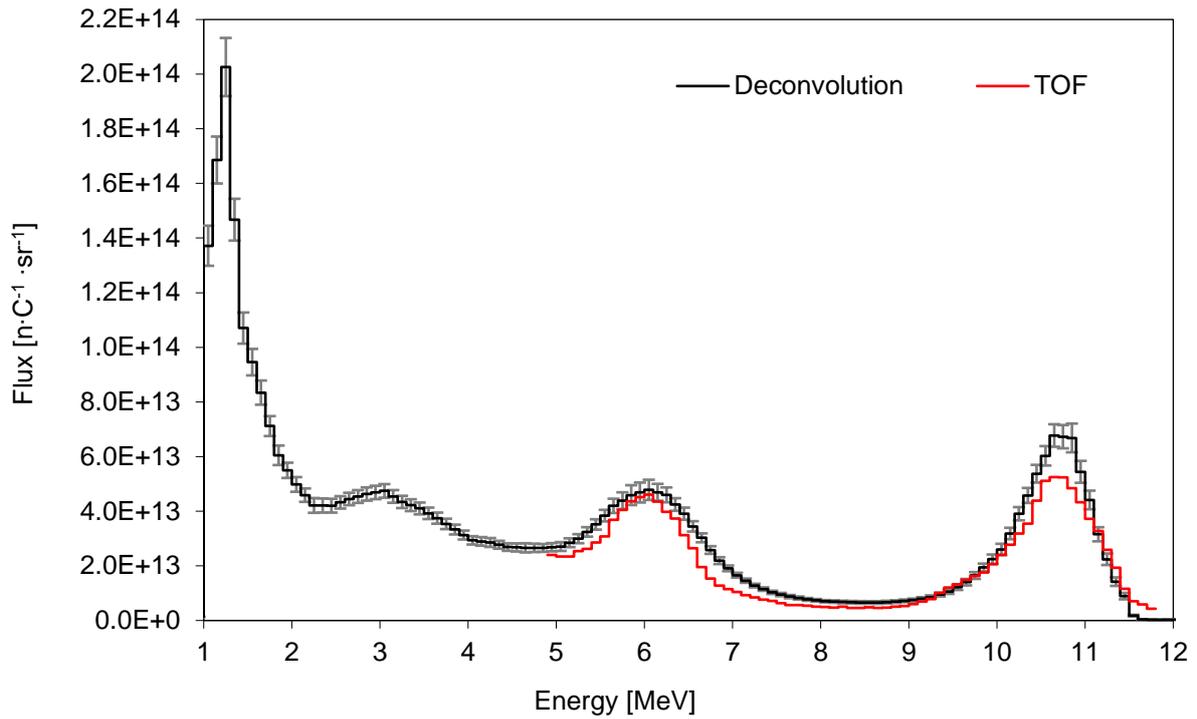

Figure 9: Comparison of deconvoluted and TOF neutron spectrum for 12.4 MeV protons.

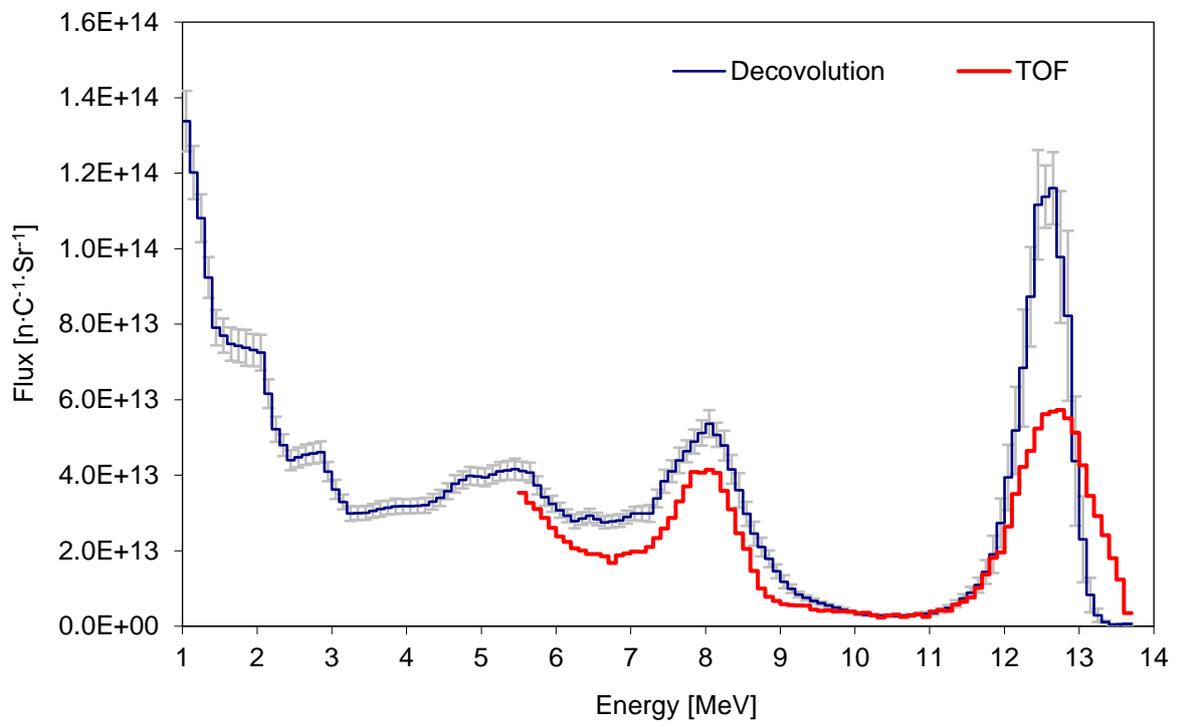

Figure 10: Comparison of deconvoluted and TOF neutron spectrum for 14.4 MeV protons.

## 4.2    Experimental reaction rates

The neutron source can be assumed as the point source, but the stack of the foils is not dimensionless. At the back of the stack, the neutron flux is decreased due to larger distance and due to neutron flux attenuation in preceding foils. To evaluate the magnitude of these effects, the thin pure $^{nat.}$Ni foils with 1.5 cm diameter and thickness of 0.01 mm were placed each 0.2 mm or behind each thick foil.

In the first part of target prior Ni, the attenuation is low (below 1.04), thus material parameter can be neglected. For suppression of uncertainties connected with measurement of monitors the $1/R^2$ law, where R is the distance from lithium target to center of actual foil, can be used (see Figure 11). In the upper parts of target namely in PTFE and Mn foils where material parameter is not negligible, the attenuation was derived from flux monitors of 0.01 mm thickness.

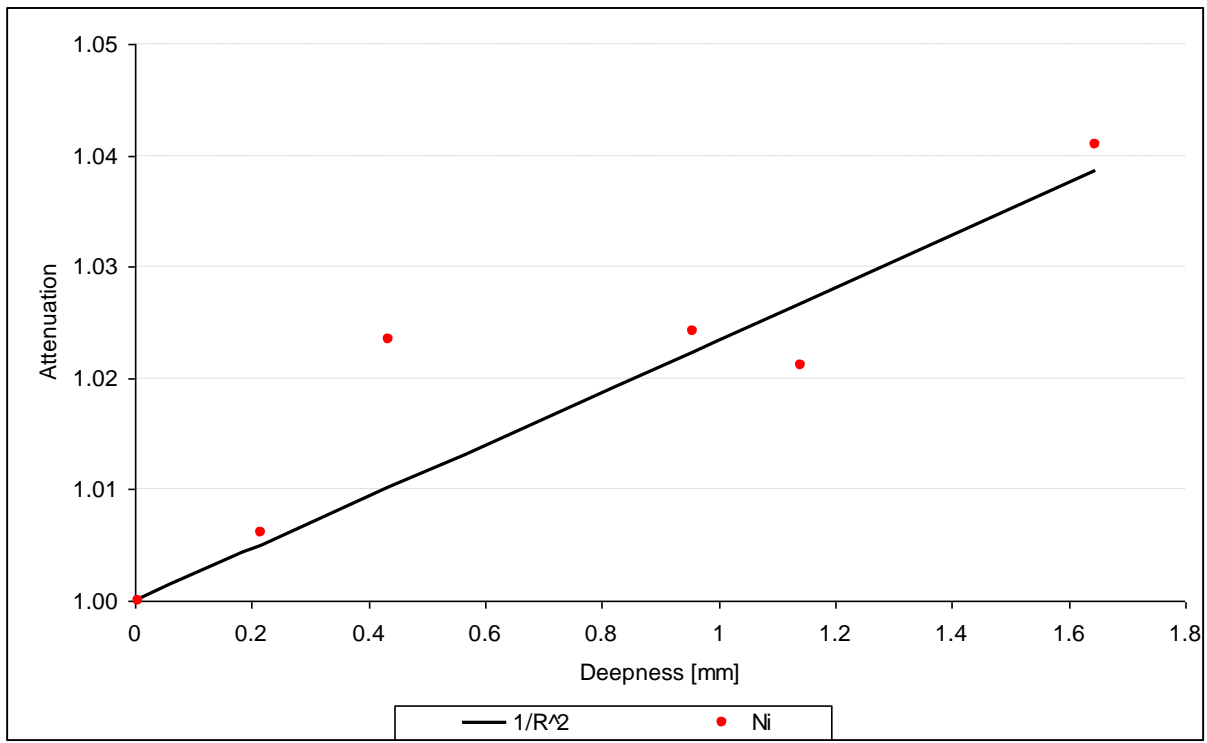

Figure 11: The theoretically predicted and measured attenuation in used target in the first target

The evaluated reaction rates were obtained from reaction rates using the geometrical attenuation factor and are listed in Table 2 for field formed by 12.4 MeV protons, and for 14.4 MeV protons in Table 3. Due to applied correction, the evaluated reaction rates are corresponding to front target position. The selection of reaction rate was done to cover the whole neutron spectrum like $^{54}$Fe(n,p) in lower part and $^{197}$Au(n,2n) in upper one. Very interesting is also covering of reactions close to threshold, namely $^{59}$Co(n,2n) in first experiment and $^{55}$Mn(n,2n) in set. Its threshold is 10.6 MeV; thus, this reaction is caused by the neutrons from the tail of the main peak.

Table 2: Evaluated reaction rates for 12.4 MeV proton beam with current 3.443 μA

| Reaction | E50% [MeV] | Reaction rate [s$^{-1}$] | Uncertainty |
|---|---|---|---|
| $^{54}$Fe(n,p) | 6.78 | 4.51E-18 | 4.1 % |
| $^{47}$Ti(n,p) | 7.43 | 9.35E-19 | 3.6 % |
| $^{46}$Ti(n,p) | 9.61 | 1.46E-18 | 8.2 % |
| $^{59}$Co(n,p) | 10.05 | 2.21E-18 | 3.9 % |
| $^{60}$Ni(n,p) | 10.25 | 4.68E-19 | 6.8 % |
| $^{56}$Fe(n,p) | 10.30 | 3.51E-19 | 3.6 % |
| $^{24}$Mg(n,p) | 10.37 | 5.81E-19 | 3.7 % |
| $^{59}$Co(n,α) | 10.43 | 7.07E-20 | 3.6 % |
| $^{48}$Ti(n,p) | 10.45 | 1.37E-19 | 3.7 % |
| $^{51}$V(n,α) | 10.57 | 1.99E-20 | 4.9 % |
| $^{197}$Au(n,2n) | 10.60 | 4.29E-18 | 3.8 % |
| $^{58}$Ni(n,x)$^{57}$Co | 10.89 | 1.89E-19 | 3.7 % |
| $^{59}$Co(n,2n) | 11.27 | 4.78E-20 | 5.7 % |
| $^{58}$Ni(n,p) | 6.71 | 5.76E-18 | 3.6 % |
| $^{27}$Al(n,α) | 10.44 | 3.66E-19 | 3.7 % |

Table 3: Evaluated reaction rates for 14.4 MeV proton beam with current 5.228 µA

| Reaction | E$_{50\%}$ [MeV] | Reaction rate [s$^{-1}$] | Uncertainty. |
|---|---|---|---|
| $^{54}$Fe(n.p) | 7.75 | 8.71E-18 | 3.6 % |
| $^{47}$Ti(n.p) | 8.37 | 1.99E-18 | 3.5 % |
| $^{46}$Ti(n.p) | 10.04 | 3.3E-18 | 3.9 % |
| $^{59}$Co(n.p) | 12.13 | 5.01E-19 | 4.4 % |
| $^{60}$Ni(n.p) | 12.24 | 1.33E-18 | 5.4 % |
| $^{54}$Fe(n,α) | 12.29 | 6.78E-19 | 5.8 % |
| $^{24}$Mg(n.p) | 12.31 | 1.55E-18 | 3.6 % |
| $^{56}$Fe(n.p) | 12.31 | 9.37E-19 | 3.5 % |
| $^{59}$Co(n.α) | 12.40 | 2.05E-19 | 3.5 % |
| $^{48}$Ti(n.p) | 12.41 | 3.99E-19 | 3.6 % |
| $^{51}$V(n.α) | 12.50 | 7.12E-20 | 4.2 % |
| $^{197}$Au(n.2n) | 12.50 | 1.1E-17 | 3.8 % |
| $^{58}$Ni(n,x)$^{57}$Co | 12.57 | 2.09E-18 | 3.5 % |
| $^{59}$Co(n.2n) | 12.58 | 2.38E-18 | 3.6 % |
| $^{19}$F(n,2n) | 12.64 | 5.72E-20 | 3.6 % |
| $^{55}$Mn(n,2n) | 12.58 | 2.56E-18 | 3.8 % |
| $^{58}$Ni(n.2n) | 12.85 | 6.17E-21 | 3.8 % |
| $^{58}$Ni(n.p) | 7.64 | 1.07E-17 | 3.5 % |
| $^{27}$Al(n.α) | 10.28 | 9.29E-19 | 3.6 % |

### 4.3   Evaluated neutron flux in target assembly

The neutron flux density in target assembly was evaluated using the spectrum measured with the scintillator detectors which was transposed to the position of the activation detectors and normalized on reaction rates of $^{27}$Al(n,α) and $^{58}$Ni(n,p). This normalization is used because the positioning of the lithium target with the accuracy of 1 mm would introduce further uncertainties in the neutron flux density. These reactions were selected because they are sensitive in different regions of neutron spectra, and their cross sections is well known (this is reflected in low uncertainties in their cross sections).

Due to relatively large volume of materials in vicinity of target assembly, the neutron spectrum in irradiated foils differs from spectrum measured far from target (see chapter 3.4). The correction factor, in form of ratios between spectrum in measuring position and spectrum in target (Figure 12) was determined by MCNP6.2 simulation. These ratios were used to transform the measured neutron spectra from the position of scintillator detectors (4 m from target) to the position of activation foils (8.6 cm from Li target).

The final neutron spectra at the position of the beginning of activation foils are tabulated in Table 4 for 12.4 MeV protons, and in Table 5 for 14.4 MeV protons.

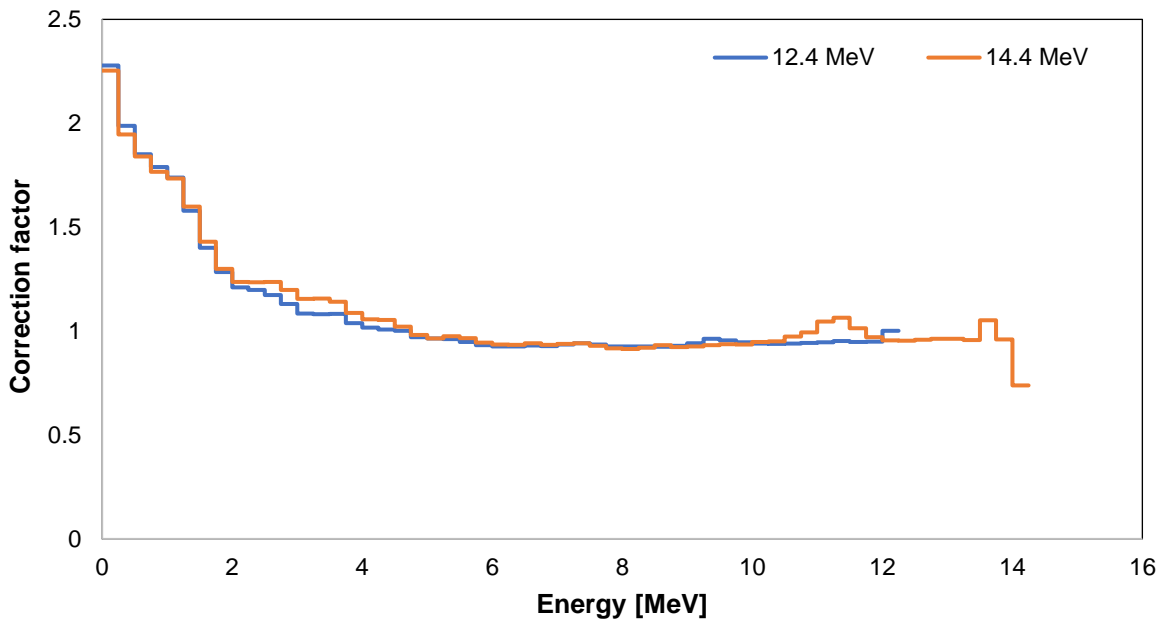

Figure 12: The theoretically predicted correction factor for target scattered neutrons

The comparison of evaluated experimental flux with calculation performed in MCNP6 (LA150H and JENDL4/HE libraries) and GEANT4 (using approach from with ENDF/B-VII.1 libraries) is plotted in Figure 13 and Figure 14. As assumed, the agreement is not very good. In case of 14.4 MeV protons, the agreement in the main peak is better than in 12.4 MeV protons irradiation. However, the second peak formed in $^7$Li(p,n)$^3$He+$^4$He reaction is practically not reflected in calculation in both energies in MCNP6 code. Geant4 reflects both peaks, but their magnitude and position differ from experiment significantly.

As many of reactions are sensitive in region of second peak as well, it can be said, the current models are not satisfactory for characterization of leakage spectra from Li target and low energy neutrons.

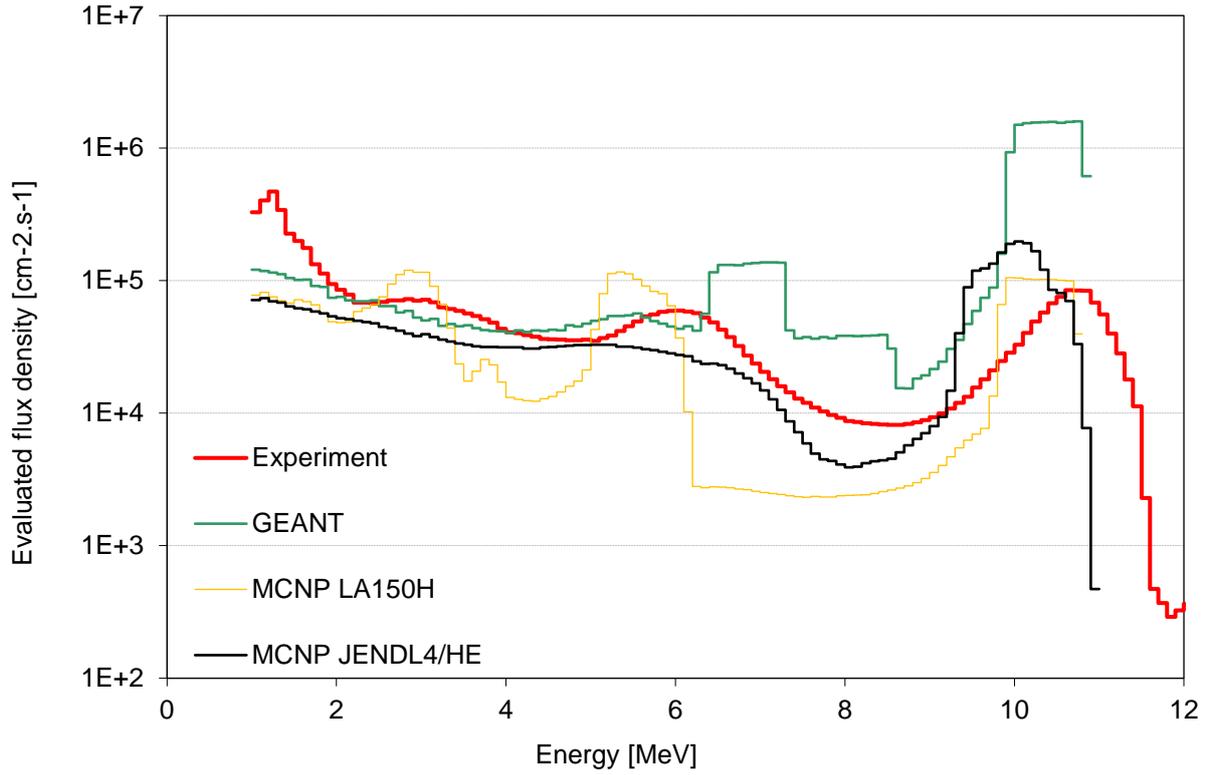

Figure 13: Calculated and evaluated neutron flux in target for 12.4 MeV protons

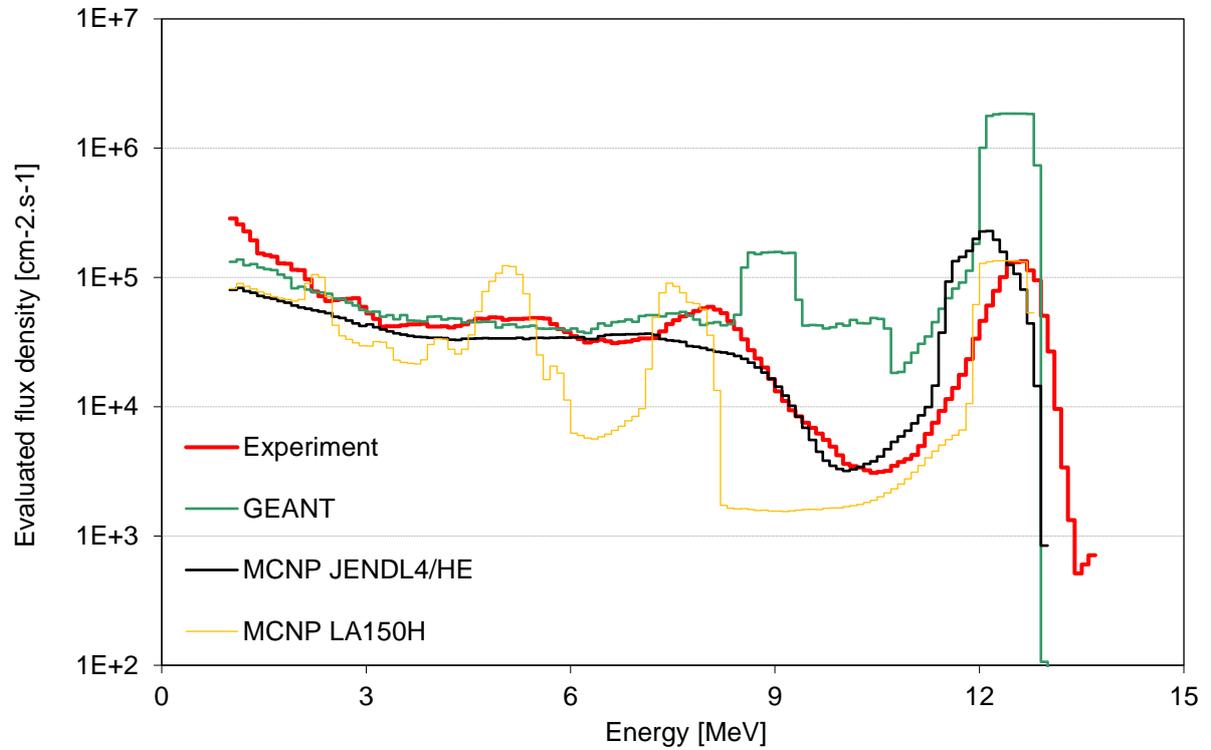

Figure 14: Calculated and evaluated neutron flux in target for 14.4 MeV protons

Table 4: Evaluated neutron spectrum in target assembly from $^{nat.}$Li target for 12.4 MeV protons

| $E_{up}$ [MeV] | Mean [cm$^{-2}$.s$^{-1}$] | Rel. unc. | $E_{up}$ [MeV] | Mean [cm$^{-2}$.s$^{-1}$] | Rel. unc. | $E_{up}$ [MeV] | Mean [cm$^{-2}$.s$^{-1}$] | Rel. unc. |
|---|---|---|---|---|---|---|---|---|
| 1 | 1.070E+4 | 12.7 % | 4.8 | 1.224E+5 | 6.5 % | 8.6 | 2.810E+4 | 7.9 % |
| 1.1 | 1.129E+6 | 6.1 % | 4.9 | 1.223E+5 | 6.2 % | 8.7 | 2.804E+4 | 8.2 % |
| 1.2 | 1.387E+6 | 5.9 % | 5.0 | 1.235E+5 | 6.4 % | 8.8 | 2.864E+4 | 7.9 % |
| 1.3 | 1.620E+6 | 6.1 % | 5.1 | 1.208E+5 | 6.7 % | 8.9 | 2.929E+4 | 7.7 % |
| 1.4 | 1.173E+6 | 6.0 % | 5.2 | 1.270E+5 | 6.4 % | 9.0 | 3.056E+4 | 7.6 % |
| 1.5 | 7.773E+5 | 6.1 % | 5.3 | 1.329E+5 | 6.3 % | 9.1 | 3.205E+4 | 7.6 % |
| 1.6 | 6.867E+5 | 5.9 % | 5.4 | 1.439E+5 | 6.1 % | 9.2 | 3.426E+4 | 7.4 % |
| 1.7 | 6.058E+5 | 6.1 % | 5.5 | 1.560E+5 | 6.2 % | 9.3 | 3.719E+4 | 7.8 % |
| 1.8 | 4.587E+5 | 5.9 % | 5.6 | 1.697E+5 | 6.3 % | 9.4 | 4.123E+4 | 8.0 % |
| 1.9 | 3.898E+5 | 6.6 % | 5.7 | 1.855E+5 | 6.9 % | 9.5 | 4.577E+4 | 8.4 % |
| 2.0 | 3.244E+5 | 5.9 % | 5.8 | 1.911E+5 | 7.6 % | 9.6 | 5.388E+4 | 8.6 % |
| 2.1 | 2.940E+5 | 6.2 % | 5.9 | 1.999E+5 | 8.5 % | 9.7 | 6.213E+4 | 8.9 % |
| 2.2 | 2.708E+5 | 6.1 % | 6.0 | 2.041E+5 | 8.4 % | 9.8 | 7.239E+4 | 8.1 % |
| 2.3 | 2.346E+5 | 7.1 % | 6.1 | 2.051E+5 | 8.3 % | 9.9 | 8.521E+4 | 8.0 % |
| 2.4 | 2.345E+5 | 6.9 % | 6.2 | 2.012E+5 | 7.2 % | 10.0 | 9.862E+4 | 8.2 % |
| 2.5 | 2.316E+5 | 6.7 % | 6.3 | 1.960E+5 | 6.6 % | 10.1 | 1.132E+5 | 8.5 % |
| 2.6 | 2.383E+5 | 6.0 % | 6.4 | 1.809E+5 | 6.2 % | 10.2 | 1.386E+5 | 7.2 % |
| 2.7 | 2.451E+5 | 5.9 % | 6.5 | 1.666E+5 | 6.4 % | 10.3 | 1.691E+5 | 6.9 % |
| 2.8 | 2.450E+5 | 6.0 % | 6.6 | 1.466E+5 | 6.3 % | 10.4 | 1.980E+5 | 6.7 % |
| 2.9 | 2.500E+5 | 6.0 % | 6.7 | 1.288E+5 | 6.2 % | 10.5 | 2.324E+5 | 6.7 % |
| 3.0 | 2.436E+5 | 6.0 % | 6.8 | 1.104E+5 | 6.2 % | 10.6 | 2.603E+5 | 6.6 % |
| 3.1 | 2.465E+5 | 6.0 % | 6.9 | 9.373E+4 | 6.5 % | 10.7 | 2.925E+5 | 6.8 % |
| 3.2 | 2.360E+5 | 5.9 % | 7.0 | 8.167E+4 | 6.2 % | 10.8 | 2.908E+5 | 7.0 % |
| 3.3 | 2.163E+5 | 6.0 % | 7.1 | 7.074E+4 | 6.2 % | 10.9 | 2.888E+5 | 8.4 % |
| 3.4 | 2.104E+5 | 6.0 % | 7.2 | 6.224E+4 | 6.1 % | 11.0 | 2.352E+5 | 8.0 % |
| 3.5 | 2.039E+5 | 6.1 % | 7.3 | 5.501E+4 | 6.1 % | 11.1 | 1.915E+5 | 8.3 % |
| 3.6 | 1.950E+5 | 6.1 % | 7.4 | 4.945E+4 | 6.2 % | 11.2 | 1.370E+5 | 8.7 % |
| 3.7 | 1.863E+5 | 6.2 % | 7.5 | 4.433E+4 | 6.7 % | 11.3 | 9.748E+4 | 9.8 % |
| 3.8 | 1.761E+5 | 6.2 % | 7.6 | 4.113E+4 | 7.0 % | 11.4 | 6.184E+4 | 11.3 % |
| 3.9 | 1.658E+5 | 6.2 % | 7.7 | 3.787E+4 | 7.5 % | 11.5 | 3.882E+4 | 13.4 % |
| 4.0 | 1.493E+5 | 6.0 % | 7.8 | 3.545E+4 | 7.3 % | 11.6 | 7.889E+3 | 17.2 % |
| 4.1 | 1.400E+5 | 6.2 % | 7.9 | 3.330E+4 | 7.2 % | 11.7 | 1.622E+3 | 21.1 % |
| 4.2 | 1.382E+5 | 6.2 % | 8.0 | 3.182E+4 | 7.0 % | 11.8 | 1.273E+3 | 21.0 % |
| 4.3 | 1.335E+5 | 6.4 % | 8.1 | 3.007E+4 | 6.9 % | 11.9 | 9.998E+2 | 21.5 % |
| 4.4 | 1.296E+5 | 6.0 % | 8.2 | 2.947E+4 | 7.2 % | 12.0 | 1.120E+3 | 21.0 % |
| 4.5 | 1.246E+5 | 6.0 % | 8.3 | 2.888E+4 | 7.6 % | 12.1 | 1.253E+3 | 21.3 % |
| 4.6 | 1.240E+5 | 6.3 % | 8.4 | 2.854E+4 | 7.6 % | 12.2 | 1.190E+3 | 21.9 % |
| 4.7 | 1.233E+5 | 7.0 % | 8.5 | 2.820E+4 | 8.1 % | | | |

Table 5: Evaluated neutron spectrum in target assembly from $^{nat.}$Li target for 14.4 MeV protons

| $E_{up}$ [MeV] | Mean [cm$^{-2}$.s$^{-1}$] | Rel. unc. | $E_{up}$ [MeV] | Mean [cm$^{-2}$.s$^{-1}$] | Rel. unc. | $E_{up}$ [MeV] | Mean [cm$^{-2}$.s$^{-1}$] | Rel. unc. |
|---|---|---|---|---|---|---|---|---|
| 1.1 | 1.494E+6 | 6.0 % | 5.3 | 2.497E+5 | 6.0 % | 9.5 | 3.937E+4 | 9.4 % |
| 1.2 | 1.342E+6 | 5.9 % | 5.4 | 2.516E+5 | 6.1 % | 9.6 | 3.584E+4 | 8.0 % |
| 1.3 | 1.184E+6 | 5.9 % | 5.5 | 2.534E+5 | 6.8 % | 9.7 | 3.240E+4 | 7.2 % |
| 1.4 | 1.012E+6 | 5.9 % | 5.6 | 2.535E+5 | 6.0 % | 9.8 | 2.893E+4 | 9.3 % |
| 1.5 | 7.998E+5 | 5.9 % | 5.7 | 2.506E+5 | 6.5 % | 9.9 | 2.565E+4 | 12.2 % |
| 1.6 | 7.782E+5 | 5.9 % | 5.8 | 2.278E+5 | 6.5 % | 10.0 | 2.208E+4 | 10.4 % |
| 1.7 | 7.557E+5 | 5.9 % | 5.9 | 2.086E+5 | 6.6 % | 10.1 | 1.893E+4 | 10.2 % |
| 1.8 | 6.717E+5 | 5.9 % | 6.0 | 1.980E+5 | 6.5 % | 10.2 | 1.819E+4 | 10.9 % |
| 1.9 | 6.667E+5 | 6.5 % | 6.1 | 1.836E+5 | 6.5 % | 10.3 | 1.773E+4 | 12.1 % |
| 2.0 | 6.010E+5 | 5.9 % | 6.2 | 1.747E+5 | 5.9 % | 10.4 | 1.690E+4 | 11.1 % |
| 2.1 | 5.953E+5 | 6.5 % | 6.3 | 1.644E+5 | 6.7 % | 10.5 | 1.609E+4 | 11.4 % |
| 2.2 | 5.060E+5 | 6.2 % | 6.4 | 1.685E+5 | 5.9 % | 10.6 | 1.639E+4 | 11.6 % |
| 2.3 | 4.076E+5 | 6.4 % | 6.5 | 1.728E+5 | 6.4 % | 10.7 | 1.665E+4 | 12.0 % |
| 2.4 | 3.746E+5 | 6.0 % | 6.6 | 1.674E+5 | 6.0 % | 10.8 | 1.825E+4 | 11.7 % |
| 2.5 | 3.435E+5 | 6.0 % | 6.7 | 1.623E+5 | 5.9 % | 10.9 | 1.958E+4 | 11.8 % |
| 2.6 | 3.492E+5 | 5.9 % | 6.8 | 1.651E+5 | 6.0 % | 11.0 | 2.057E+4 | 13.0 % |
| 2.7 | 3.548E+5 | 5.9 % | 6.9 | 1.665E+5 | 6.0 % | 11.1 | 2.210E+4 | 15.4 % |
| 2.8 | 3.579E+5 | 5.9 % | 7.0 | 1.719E+5 | 6.0 % | 11.2 | 2.596E+4 | 14.8 % |
| 2.9 | 3.606E+5 | 6.1 % | 7.1 | 1.762E+5 | 6.0 % | 11.3 | 3.215E+4 | 15.2 % |
| 3.0 | 3.099E+5 | 6.2 % | 7.2 | 1.763E+5 | 6.6 % | 11.4 | 3.944E+4 | 15.5 % |
| 3.1 | 2.743E+5 | 7.0 % | 7.3 | 1.771E+5 | 7.5 % | 11.5 | 4.850E+4 | 15.9 % |
| 3.2 | 2.493E+5 | 5.9 % | 7.4 | 2.006E+5 | 7.0 % | 11.6 | 6.004E+4 | 16.8 % |
| 3.3 | 2.181E+5 | 6.5 % | 7.5 | 2.275E+5 | 8.0 % | 11.7 | 7.324E+4 | 18.2 % |
| 3.4 | 2.187E+5 | 5.9 % | 7.6 | 2.442E+5 | 7.1 % | 11.8 | 9.211E+4 | 21.9 % |
| 3.5 | 2.193E+5 | 6.1 % | 7.7 | 2.614E+5 | 7.7 % | 11.9 | 1.220E+5 | 25.9 % |
| 3.6 | 2.232E+5 | 6.0 % | 7.8 | 2.721E+5 | 6.3 % | 12.0 | 1.751E+5 | 23.8 % |
| 3.7 | 2.272E+5 | 6.6 % | 7.9 | 2.872E+5 | 6.5 % | 12.1 | 2.416E+5 | 22.0 % |
| 3.8 | 2.268E+5 | 5.9 % | 8.0 | 3.007E+5 | 6.5 % | 12.2 | 3.179E+5 | 22.3 % |
| 3.9 | 2.293E+5 | 6.4 % | 8.1 | 3.108E+5 | 6.7 % | 12.3 | 4.129E+5 | 22.7 % |
| 4.0 | 2.183E+5 | 6.0 % | 8.2 | 2.936E+5 | 6.3 % | 12.4 | 5.267E+5 | 15.1 % |
| 4.1 | 2.184E+5 | 5.9 % | 8.3 | 2.765E+5 | 8.3 % | 12.5 | 6.738E+5 | 13.0 % |
| 4.2 | 2.192E+5 | 5.9 % | 8.4 | 2.403E+5 | 9.1 % | 12.6 | 6.858E+5 | 7.3 % |
| 4.3 | 2.138E+5 | 6.3 % | 8.5 | 2.082E+5 | 12.6 % | 12.7 | 6.992E+5 | 8.3 % |
| 4.4 | 2.202E+5 | 6.3 % | 8.6 | 1.732E+5 | 10.8 % | 12.8 | 5.925E+5 | 17.9 % |
| 4.5 | 2.261E+5 | 6.3 % | 8.7 | 1.426E+5 | 13.0 % | 12.9 | 4.982E+5 | 27.4 % |
| 4.6 | 2.381E+5 | 5.9 % | 8.8 | 1.235E+5 | 11.9 % | 13.0 | 2.650E+5 | 39.2 % |
| 4.7 | 2.510E+5 | 6.1 % | 8.9 | 1.053E+5 | 12.4 % | 13.1 | 1.401E+5 | 49.4 % |
| 4.8 | 2.503E+5 | 5.9 % | 9.0 | 8.561E+4 | 12.7 % | 13.2 | 5.030E+4 | 55.0 % |
| 4.9 | 2.573E+5 | 5.9 % | 9.1 | 6.870E+4 | 14.8 % | 13.3 | 1.771E+4 | 60.7 % |
| 5.0 | 2.561E+5 | 5.9 % | 9.2 | 5.804E+4 | 11.4 % | 13.4 | 6.952E+3 | 34.1 % |
| 5.1 | 2.449E+5 | 6.0 % | 9.3 | 4.908E+4 | 8.6 % | 13.5 | 2.684E+3 | 39.7 % |
| 5.2 | 2.496E+5 | 6.0 % | 9.4 | 4.400E+4 | 8.6 % | 13.6 | 3.151E+3 | 11.2 % |

## 4.4 Cross section validation using measured reaction rates

The measured reaction rates were compared with the values obtained by folding of the evaluated spectra (described above in chapter 4.3) with the cross sections from the selected libraries. As the spectra are well defined, with low uncertainties, the reaction rates derived by folding of selected cross section with spectrum has also small uncertainties and this comparison can be understood as a validation. The validation was performed for IRDFF-II [35], JEFF-3.3 [36], JENDL-4 [37] and ENDF/B-VIII [24] nuclear data libraries.

Due to applied flux loss correction, which corresponds to $1/R^2$ law, the flux in each foil is identical and the flux in beginning of target listed in Table 4 or Table 5 is used. It is worth noting, the best agreement was reached for IRDFF-II nuclear data library.

Table 6: Validation of selected reaction rates in $^7$Li(p,n) field from 12.4 MeV protons by means of C/E-1 comparison

| Reaction | $E_{50\%}$ [MeV] | IRDFF-II | JEFF-3.3 | JENDL-4 | ENDF/B-VIII | Unc. |
|---|---|---|---|---|---|---|
| $^{54}$Fe(n,p) | 6.78 | 1.9 % | -2.3 % | 5.2 % | 2.0 % | 4.2 % |
| $^{47}$Ti(n,p) | 7.43 | 7.8 % | 8.4 % | 11.7 % | 23.6 % | 3.7 % |
| $^{46}$Ti(n,p) | 9.61 | 11.7 % | 12.2 % | 9.8 % | 4.8 % | 8.3 % |
| $^{59}$Co(n,p) | 10.05 | 3.2 % | -7.1 % | -5.5 % | 3.2 % | 4.1 % |
| $^{60}$Ni(n,p) | 10.25 | 18.1 % | 18.8 % | 20.9 % | 18.1 % | 6.9 % |
| $^{56}$Fe(n,p) | 10.30 | -2.5 % | -4.0 % | 0.2 % | -2.5 % | 3.9 % |
| $^{24}$Mg(n,p) | 10.37 | 3.7 % | 8.4 % | 10.3 % | 7.7 % | 4.1 % |
| $^{59}$Co(n,$\alpha$) | 10.43 | -1.9 % | -4.6 % | 1.0 % | 1.7 % | 4.0 % |
| $^{48}$Ti(n,p) | 10.45 | -0.5 % | 0.1 % | -1.9 % | 10.6 % | 4.1 % |
| $^{51}$V(n,$\alpha$) | 10.57 | -1.7 % | -1.2 % | 8.4 % | 12.2 % | 5.3 % |
| $^{197}$Au(n,2n) | 10.60 | -3.4 % | -5.5 % | 22.6 % | -6.1 % | 4.3 % |
| $^{58}$Ni(n,x)$^{57}$Co | 10.89 | - | -20.9 % | -10.6 % | -18.1 % | 3.7 % |
| $^{59}$Co(n,2n) | 11.27 | -2.7 % | -2.7 % | 11.7 % | 14.0 % | 6.2 % |

Table 7: Validation of selected reaction rates in $^7$Li(p,n) field from 14.4 MeV protons by means of C/E-1 comparison

| Reaction | $E_{50\%}$ [MeV] | IRDFF-II | JEFF-3.3 | JENDL-4 | ENDF/B-VIII | Unc. |
|---|---|---|---|---|---|---|
| $^{54}$Fe(n,p) | 7.75 | -1.2 % | -3.0 % | 0.6 % | -1.0 % | 3.9 % |
| $^{47}$Ti(n,p) | 8.37 | 1.3 % | 1.3 % | 2.1 % | 15.9 % | 6.8 % |
| $^{46}$Ti(n,p) | 10.04 | 7.8 % | 7.8 % | 2.9 % | 4.2 % | 4.1 % |
| $^{59}$Co(n,p) | 12.13 | 7.2 % | -1.7 % | -1.3 % | 4.8 % | 4.6 % |
| $^{60}$Ni(n,p) | 12.24 | 3.8 % | 1.2 % | 7.9 % | 9.8 % | 5.3 % |
| $^{54}$Fe(n,$\alpha$) | 12.29 | -5.4 % | -17.4 % | -5.5 % | -61.7 % | 6.4 % |
| $^{24}$Mg(n,p) | 12.31 | 3.0 % | 10.1 % | 10.1 % | 10.1 % | 5.0 % |
| $^{56}$Fe(n,p) | 12.31 | -1.7 % | -2.5 % | -3.9 % | -1.6 % | 5.0 % |
| $^{59}$Co(n,$\alpha$) | 12.40 | -2.9 % | -5.0 % | -3.2 % | -0.9 % | 5.3 % |
| $^{48}$Ti(n,p) | 12.41 | 2.2 % | 2.3 % | -3.9 % | 0.5 % | 5.4 % |
| $^{51}$V(n,$\alpha$) | 12.50 | -4.8 % | -5.0 % | -0.4 % | 1.5 % | 6.3 % |
| $^{197}$Au(n,2n) | 12.50 | 2.9 % | 4.2 % | 16.0 % | 4.2 % | 6.2 % |

| | | | | | |
|---|---|---|---|---|---|
| $^{58}$Ni(n,x)$^{57}$Co | 12.57 | - | -7.7 % | 2.5 % | -10.2 % | 6.6 % |
| $^{59}$Co(n,2n) | 12.58 | -7.2 % | -6.4 % | -13.5 % | -4.0 % | 6.7 % |
| $^{19}$F(n,2n) | 12.64 | 10.7 % | 40.1 % | 75.7 % | 40.1 % | 7.3 % |
| $^{55}$Mn(n,2n) | 12.58 | -2.8 % | 4.9 % | -1.0 % | -3.9 % | 6.8 % |

### 4.5    $^7$Be production in the lithium target and forward directed peak neutrons

The cross section for the $^7$Be production by protons in the lithium target - $\sigma(^7$Be) - and the number of the forward directed peak neutrons - $(d\sigma/d\Omega)_{\theta=0}$ - are routinely measured and tabulated quantities for the p+$^7$Li based quasi monoenergetic neutron sources [38]. These two quantities are connected through the Uwamino's "index of forwardness" – R - [29], valid in the proton energy region 20-40 MeV, and Tadeucci systematics above 80 MeV. Below 20 MeV, the experiments with the measured values of the $^7$Be production and the forward directed peak neutrons do not exist.

After both irradiations, the lithium target was extracted from the target station and the activity of the produced $^7$Be was measured using the HPGe detector (intensities in Tab. 1). The number of the forward directed peak neutrons was measured by the scintillator detector (the values from deconvolution and TOF differ for approx. 20%, TOF values are shown as in [38] ). Further details of the procedure are described in [38] The measured values together with the "index of forwardness" – R - are shown in Tab. 8.

Table 8: The experimentally measured data. The quoted energy is the average proton energy in 0.5-mm thick lithium target.

| Energy (MeV) | R(sr$^{-1}$) | $\sigma(^7$Be) | $(d\sigma/d\Omega)_{\theta=0}$ (mb/sr) |
|---|---|---|---|
| 12.4 | 0.10(0.01) | 51.2(3.0) | 5.3(0.7) |
| 14.4 | 0.13(0.01) | 47.7(3.0) | 6.6(0.7) |

### 5    Conclusions

The developed methodology of testing the evaluated nuclear data by well- defined quasi mono energetic neutron field from $^7$Li(p,n) reaction seems to be a robust testing method sensitive in energy regions different from commonly used fission neutron formed fields. This is promising approach for testing of data sets for use in neutron dosimetry especially in accelerator applications or even space applications.

The methodology was tested on a large set of common dosimetry reactions in IRDFF-II, ENDF/B-VIII, JEFF-3.3 and JENDL-4 reaction rates. In common dosimetry reactions, clearly the best agreement was reached in IRDFF-II data set. The new recommended dosimetry reaction $^{58}$Ni(n,x)$^{57}$Co shows better agreement in the case of JENDL-4, thus it can be noted that this evaluation seems to be the most realistic in the lower energy region.

The calculation realized with MCNP6.2 and its comparison with measured spectra can lead to conclusion, that none of the used approaches can successfully reproduce the measured spectrum.

Acknowledgments

The presented work has been realized within Institutional Support by the Ministry of Industry and Trade and with the use of the infrastructure Reactors LVR-15 and LR-0, which is financially supported by the Ministry of Education, Youth and Sports – project LM2015074, the SANDA project funded under H2020-EURATOM-1.1 contract 847552.
The measurements were carried out at the CANAM Research Infrastructure of the NPI CAS. Computational resources were supplied by the project "e-Infrastruktura CZ" (e-INFRA CZ LM2018140 ) supported by the Ministry of Education, Youth and Sports of the Czech Republic.